\documentclass[universe,article,submit,moreauthors,pdftex]{Definitions/mdpi} 

\firstpage{1} 
\articlenumber{}
\pubvolume{1}
\issuenum{1}
\articlenumber{0}
\pubyear{2021}
\copyrightyear{2021}
\externaleditor{Academic Editor: }
\datereceived{} 
\dateaccepted{}
\datepublished{} 
\hreflink{https://doi.org/} 

\usepackage{graphicx}
\usepackage{amssymb}
\preto{\abstractkeywords}{\nolinenumbers}
\usepackage{enumitem}
\usepackage{amsmath}
\usepackage[normalem]{ulem}

\def\al{\alpha}
\def\be{\beta}

\def\de{\delta}
\def\ep{\epsilon}
\def\ze{\zeta}
\def\et{\eta}
\def\th{\theta}
\def\vth{\vartheta}
\def\ka{\kappa}
\def\la{\lambda}
\def\rh{\rho}

\def\si{\sigma}

\def\ta{\tau}

\def\ph{\phi}
\def\vph{\varphi}

\def\ps{\psi}
\def\om{\omega}

\def\Si{\Sigma}

\def\Ph{\Phi}

\def\cL{{\cal L}}

\def\mn{{\mu\nu}}

\def\prt{\partial}

\def\pt#1{\phantom{#1}}

\newcommand{\beq}{\begin{equation}}
\newcommand{\eeq}{\end{equation}}
\newcommand{\bea}{\begin{eqnarray}}
\newcommand{\eea}{\end{eqnarray}}
\newcommand{\rf}[1]{(\ref{#1})}

\def\st{\tilde{s}}

\def\fr#1#2{{{#1} \over {#2}}}

\def\frac#1#2{{\textstyle{{#1}\over {#2}}}}

\def\rf#1{(\ref{#1})}

\usepackage{xcolor}

\Title{Analysis of birefringence and dispersion effects from spacetime-symmetry breaking in gravitational waves}

\Author{Kellie O'Neal-Ault $^{1}$, Quentin G.\ Bailey $^{1}$, Tyann Dumerchat$^{2}$, Le\"ila Haegel $^{2}$, Jay Tasson $^{3}$}
\address{%
$^{1}$ \quad Embry-Riddle Aeronautical University, Prescott, AZ, 86301, USA\\
$^{2}$ \quad Universit\'{e} de Paris, CNRS, Astroparticule et Cosmologie,  F-75006 Paris, France\\
$^{3}$ \quad Carleton College, Northfield, MN 55057, USA}

\AuthorNames{Kellie Ault-O'Neal, Quentin G.\ Bailey, Tyann Dumerchat, Le\"ila Haegel, Jay Tasson}

\corres{Correspondence: baileyq@erau.edu; Tel.: 928-777-3932}

\abstract{
In this work, we review the effective field theory framework to search for Lorentz and CPT symmetry breaking during the propagation of gravitational waves.  
The article is written so as to bridge the gap between the theory of spacetime-symmetry breaking and the analysis of  gravitational-waves signals detected by ground-based interferometers.
The primary physical effects beyond General Relativity that we explore here are dispersion and birefringence of gravitational waves.
We discuss their implementation in the open-source LIGO-Virgo algorithm library suite, as well as the statistical method used to perform a Bayesian inference of the posterior probability of the coefficients for symmetry-breaking.
We present preliminary results of this work in the form of simulations of modified gravitational waveforms, together with sensitivity studies of the measurements of the coefficients for Lorentz and CPT violation. 
The findings show the high potential of gravitational wave sources across the sky to probe sensitively for these signals of new physics.}


\begin{document}

\section{Introduction}

Gravitational waves (GWs) are now a ripe testing ground for many aspects of gravitational physics \cite{PhysRevLett.116.061102,LIGOScientific:2016lio,LIGOScientific:2019fpa,Abbott:2020jks}.
One of the principle foundations of General Relativity is the Einstein
Equivalence Principle, which includes the universality of freefall 
and the spacetime-symmetry principle of the local Lorentz invariance of physics \cite{Will:2014kxa}.
The latter principle has seen a boom in tests in the last 20+ years \cite{datatables}, owing primarily to an interesting piece of motivation: that in a fundamental unified theory of physics, local Lorentz invariance may be broken
\cite{ksstring89,gp99,chkl01}.  
The development of an effective field theory framework that describes spacetime-symmetry violations makes comparisons between vastly different kinds of tests possible, 
generalizing older kinematical test frameworks with a modern viewpoint \cite{ck97,ck98,k04}.

The specific consequences of local Lorentz-symmetry breaking for GWs has been studied in several works,  
within a general effective field theory framework \cite{bk06,km16,km18,Xu:2019fyt,Xu:2021dcw,nascimento21}, 
and in specific models \cite{yunes16,Berti:2018cxi,Amarilo:2019lfq,ferrari07,tso16,Wang_2020,Qiao_2019}. 
In particular, 
the effects on propagation have been determined for generic Lorentz-violating terms in the linearized gravity limit \cite{Mewes:2019}, 
which is the focus in this work.  

Examples of searches for Lorentz violation in gravity include table-top tests like gravimetry \cite{Muller:2007es,Chung:2009rm,Hohensee:2011wt,Hohensee:2013hra,Flowers:2016ctv,Shao:2017bgz,Ivanov:2019ouz}, short-range gravity tests \cite{Long:2014swa,Shao:2016cjk,Shao:2018lsx}, near-Earth tests \cite{Bourgoin:2016ynf,Bourgoin:2017fpo,Bourgoin:2020ckq,Bars:2019lek}, solar system planetary tests \cite{Iorio:2012gr,Hees:2013iv,Poncin-Lafitte:2016nqd}, 
and astrophysical tests with pulsars \cite{Shao:2014oha,Shao:2018vul,Wex:2020ald}.
Measurements of simultaneous gravitational and electromagnetic radiation have yielded limits on certain types of Lorentz violation in gravity versus light \cite{Abbott_2017}.
Recent work has begun to look at the available GW catalog to place constraints on coefficients describing Lorentz violation for gravity \cite{Liu:2020slm, shao20, wang21}.
Additionally,  
closely-related searches for parameterizations of deviations from General Relativity have been completed 
\cite{LIGOScientific:2019fpa,Abbott:2020jks,Wang_2020,Wang:2020cub}.

In this article, we discuss the derivation of the polarization-dependent dispersion of GW due to Lorentz and Charge-Parity-Time reversal (CPT) symmetry breaking.
We describe the implementation of the modified GW strain in the LIGO-Virgo \cite{LIGOScientific:2016lio, LIGOScientific:2019fpa}
algorithm library, 
\texttt{LALSuite} \cite{lalsuite},
as well as the statistical method used to infer the posterior probability of the coefficients for symmetry-breaking.
In order to link the theoretical derivation to the analysis of astrophysical signals, we provide detailed explanations of the steps necessary to measure the coefficients for CPT and Lorentz violation, alongside simulations of the modified signals and studies of the sensitivity of current GW interferometers for parameter inference.

The layout of the article is as follows.  In Section \ref{theory},  we describe the theoretical methodology for effective field theory descriptions of local Lorentz violation, including a scalar field example, and an effective quadratic action for the spacetime metric fluctuations.
This Section also includes the discussion of the modified plane wave solutions, and the conversion of various expressions 
to Syst\`eme International (SI) units.
Following this, 
in Section \ref{lalsuite} we describe the implementation of the modified GW signals and the statistical method used for the inference of the coefficients controlling the Lorentz and CPT-breaking effects on propagation.
Section \ref{simulation} includes simulations for a particular subset of the possible forms of Lorentz and CPT violations.
A summary and outlook is included in Section \ref{conclusion}.

For the bulk of the paper, we work in natural units where $\hbar=c=1$
and Newton's gravitational constant is $G_N \neq 1$, except when we explicitly write some expressions in SI units.
Our convention is to work with Greek letters for spacetime indices and latin letters for spatial indices.
The flat spacetime metric signature aligns with the common General Relativity related convention $-+++$.

\section{Theoretical Framework}
\label{theory}

\subsection{Background and General Relativity}
\label{background}

As in the typical gravitational wave scenario, 
we expand the spacetime metric $g_\mn$ around
a flat Minkowski background $\et_\mn$ as
\beq
g_{\mu \nu} = \eta_{\mu \nu}+h_{\mu \nu}.
\label{metric}
\eeq
Far from the source at the detectors, GWs are treated as perturbations $h_\mn$ around the Minkowski metric 
where ${h_{\mu\nu} << \eta_{\mu\nu}}$ (e.g., components on the order of ~$10^{-21}$).
However, 
one does not assume $h_\mn$ is small compared to unity in all regions.
In particular, 
in solving for the complete solution in the far radiation zone, 
one needs to solve in the near zone as well, 
for example in a post-Newtonian series \cite{pw00,pw02}.

In standard General Relativity, one solves the Einstein field equations for the metric; The form in \rf{metric} is a rewritten form, not yet a specific solution. The full Einstein field equations can be written in the ``relaxed" form, 
as
\beq
(G_L)^\mn =\ka [ (T_M)^\mn +\ta^\mn ],
\label{relaxed}
\eeq
where $(T_M)^\mn$ is the matter stress-energy tensor, 
and $\ta^\mn$ is the energy-momentum pseudo tensor \cite{pw14}, 
and $\ka=8 \pi G_N$.
Note that in this equation, 
$(G_L)^\mn$ is the
Einstein tensor linearized in $h_\mn$.

In the wave zone, 
where the gravitational fields are weak, 
the equation \rf{relaxed} becomes simply the ``vacuum" equations $(G_L)^\mn =0$, 
which admit wave solutions with two transverse degrees of freedom after choosing
a gauge.
The transverse-traceless gauge (TT-gauge) is used to describe the propagation of GWs; 
in this gauge, GR predicts two linearly independent polarizations labelled ``$+$'' and ``$\times$'', with a phase angle difference of $\pi/4$, 
\begin{equation}
h_{\mu\nu} = \begin{pmatrix}
 0 & 0 & 0 & 0\\
 0 & h_{+} & h_{\times} & 0\\
 0 & h_{\times} & -h_{+} & 0\\
 0 & 0 & 0 & 0
\end{pmatrix}.
\label{TT}
 \end{equation} 
The observable signal comes from the LIGO and Virgo detector responses to the incoming GW, 
\beq
    S_{A}(t,\theta,\phi,\psi) = F_{A,+}(\theta,\phi,\psi)\,h_{+}(t,\theta,\phi,\psi,\tau)\,+\,F_{A,\times}(\theta,\phi,\psi)\,h_{\times}(t,\theta,\phi,\psi,\tau),
    \label{antenna}
\eeq
where $F_{A,*}$ are the antenna response patterns of the detectors. 
The angles $\theta$ and $\phi$ are the source sky locations, 
$\tau$ is the time delay between detectors receiving the signal, 
and $\psi$ is the GW frame rotation with respect to the detectors' frame. Note the individual, 
polarization terms above are not gauge independent, 
as they depend on $\psi$, 
yet the entire observed signal is gauge independent.
We return to this point below when discussing Lorentz violation effects
for GWs.

\subsection{Spacetime-symmetry breaking scenario}
\label{spacetime}

We consider the observable effects on gravitational wave propagation subject 
to Lorentz- and CPT-breaking terms in an effective field theory framework
known as the Standard-Model Extension (SME) \cite{ck97,ck98,k04,bk06,kt11,bkx15,km16}. 

\subsubsection{Scalar field example}
\label{scalar}

To help understand this framework, 
and how the action is developed,
we consider first a scalar field action
in flat spacetime.
A free\, massless scalar field $\Ph$ \, is described by the action 
\beq
    I_{\rm sc}= -\frac 12 \int d^4x \, \et^\mn (\prt_\mu \Ph) (\prt_\nu \Ph ).
    \label{scalar1}
\eeq
When varying this action $\Ph \rightarrow \Ph + \de \Ph$ one obtains, 
to first order in $\de \Ph$, and applying the Leibniz rule,
\bea
    \de I_{\rm sc} &=& -\int d^4x \, \et^\mn (\prt_\mu \de \Ph ) (\prt_\nu \Ph )
    \nonumber\\
    &=& -\int d^4x \, [ \prt_\mu \left( \de \Ph  (\prt^\mu \Ph ) \right) 
    - \de \Ph \prt_\mu \prt^\mu \Ph ].
    \label{scalar2}
\eea
Because of the total derivative,
the first term is total 4 divergence and hence is 
normally considered a surface term, 
to be evaluated on the 3 dimensional hypersurface $\Si$ bounding the volume of spacetime considered.
Since the variational principle in field theory normally assumes the 
variation $\de \Ph$ vanishes on the boundary, 
this term vanishes.
What is left is proportional to the arbitrary variation $\de \Ph$
and therefore if $\de I = 0 $ is imposed we obtain the field equations:
\beq
\Box \Ph = 0,
\label{scalareq1}
\eeq
where $\Box=\prt_\alpha \prt^\alpha$.

In the effective field theory framework description of Lorentz violation, 
terms are added to the action \rf{scalar1} that are formed
from contractions of general background coefficients with arbitrary numbers of indices $k^{\mu\nu\la...}$
and terms involving the scalar field like $\prt_\mu \Ph \prt_\nu \Ph$.
This is based on the premise that any form of Lorentz violation can be described
by the coupling of known matter fields to a fixed background field $k^{\mu\nu\la...}$ \cite{ck97,ck98}.
Under {\it particle} Lorentz transformations, 
the matter fields transform as tensors, while the background field remain fixed.
On the other hand,
under {\it observer} transformations, 
both background and matter fields transform.
The latter condition reflects the idea that physics should be independent 
of coordinates.  
These concepts are detailed in the literature.
Most notably, 
see references \cite{Bertschinger:2013jla,bertschinger19} for illustrations in classical mechanics contexts.

There are several treatments of the origin of Lorentz violation that can play a role in 
the phenomenology of the effective field theory test framework (SME). 
The Lorentz violation can be explicit, 
in which the coefficients are prescribed, 
{\it a priori} unknown background fields, 
unaccompanied by additional dynamical modes.
On the other hand, a more elegant mechanism of 
spontaneous Lorentz-symmetry breaking can be considered.
In this latter case,
the underlying action for the model is Lorentz invariant but through a dynamical process, nonzero vacuum expectation values for tensor fields can arise \cite{ksstring89}.
Other scenarios with alternative geometries like Riemann-Finsler geometry 
have been explored \cite{Kostelecky:2011qz,Lammerzahl:2012kw,AlanKostelecky:2012yjr,Javaloyes:2013ika,Schreck:2014hga,SILVA201474}.  
Much theoretical discussion of these topics exists in the literature \cite{k04,PhysRevD.82.044020,Arraut:2015nqa,bk05,bkx08,bluhm15,kostelecky:2021xhb}, 
but we do not delve into details here.

As a simple start, one
might consider trying to add a vector coupled to a first derivative of the scalar to \rf{scalar1}, 
as in 
\beq
    \Delta I_{\rm sc}=\int d^4x \, k_\nu ( \Ph \partial^{\nu}\Ph ) 
    \label{scalarLV1}
\eeq
for arbitrary background vector $k_\nu$ (we assume
the explicit symmetry breaking case for the moment). 
However, 
this can be shown to be equivalent to a surface term:
\bea
    \Delta I_{\rm sc}&=&\frac{1}{2}\int d^4x \, k_{\nu}\partial^{\nu}(\Ph^2) \nonumber\\
    &=&\frac{1}{2}\int d^3 \Si_\nu \, k^{\nu}\,\Ph^2 ,
    \label{surfacek}
\eea
where $d^3\Si_\nu$ is the hypersurface ``area" element. 
Since the variation $\de \Ph$ is assumed to vanish on the hypersurface, 
this contribution will vanish from the field equations.
Alternatively, 
variation of \rf{scalarLV1} yields a null result more directly:
\bea
    \delta \Delta I_{\rm sc} &=& \int d^4x \, k_{\nu}\left[\delta \Ph \, \partial^{\nu}\Ph +\Ph \,\delta(\partial^{\nu}\Ph)\right] 
    \nonumber\\
    &=&\int d^4x \, k_{\nu}(\partial^{\nu}\Ph - \partial^{\nu}\Ph )\, \delta \Ph,
\eea
where the last line identically is zero.

To obtain Lorentz-violating terms that yield physical results
we modify the action in \rf{scalar1} as
\beq
    I_{\rm sc}= -\frac 12 \int d^4x \, \left( \et^\mn (\prt_\mu \Ph ) (\prt_\nu \Ph ) 
    +  (\prt_\mu \Ph ) k^\mn \prt_\nu \Ph ) \right),
    \label{scalarLV2}
\eeq
where $k^\mn$ are the coefficients for Lorentz violation \cite{ck98,Edwards:2018lsn}, 
containing $10$ independent coefficients describing 
the degree of Lorentz violation.
Note that we assume here that the coefficients are constants
in the chosen coordinate system (i.e., the partials vanish, $\prt_\al k^\mn = 0$).
Upon variation, 
as in \rf{scalar1}, 
we obtain the modified field equations
\beq
\Box \Ph + k^\mn \prt_\mu \prt_\nu \Ph=0.
\label{scalareqn2}
\eeq

To complete the discussion here we also consider
the plane wave solutions to \rf{scalareqn2}.
This is achieved by assuming $\Ph$ takes 
the form $\Ph = A e^{i p_\mu x^\mu}$, 
where $x^\mu$ is spacetime position and 
$p^\mu = (\om, \vec p )$ is the 
four-momentum for the plane wave.
This yields the momentum-space equation
\beq
p_\mu p^\mu + k_\mn p^\mu p^\nu = 0.
\label{disp1}
\eeq
Using the definition of the four-momentum 
we can write this out in a space and time decomposed form:
\beq
\om^2 (1-k_{00}) -2 k_{0j} p^j \om  - k_{ij} p^i p^j - \vec p^2 = 0.
\label{disp2}
\eeq
We can solve for the dispersion relation $\om (\vec p)$
and then expand the result to leading order in 
the coefficients $k_\mn$.
We obtain
\beq
\om \approx |\vec p| 
\left( 1+ \frac 12 (k_{00} + 2 k_{0j} {\hat p}^j + k_{ij} {\hat p}^i {\hat p}^j ) \right)
\label{disp3}.
\eeq

This dispersion would modify the propagation of the scalar mode, 
in particular its speed $v=\om/|\vec p|$ can be written as
\beq
v \approx 1+ \frac 12 (k_{00} + 2 k_{0j} {\hat p}^j + k_{ij} {\hat p}^i {\hat p}^j ).
\label{speed}
\eeq
Note the directional dependence of the speed due to the anisotropic coefficients
$k_{0j}$ and $k_{ij}$.
Even in the case of the isotropic limit, 
where only $k_{00}$ appears,
due to the observer Lorentz covariance, 
this limit is 
is a special feature of a particular observer frame.
For example when viewed by an observer boosted by small $\be^j$, 
anisotropic terms will arise ( e.g., $(k^\prime)_{0j} \sim -\be^j k_{00}$ ).

In the typical effective field theory treatment of 
searches for Lorentz violation, 
additional, 
``higher order", 
terms are also included 
\cite{km09}.
Thus the Lagrange density
takes the form
\beq
    I_{\rm sc} = -\frac 12 \int d^4x \, \left( \et^\mn (\prt_\mu \Ph ) (\prt_\nu \Ph ) 
    +  (\prt_\mu \Ph ) \sum_d (k^{(d)})^{\mu\nu\la...} (\prt_\nu \prt_\la ...\Ph ) \right),
    \label{scalarLV3}
\eeq
where now the coefficients are labeled $d$ for the mass dimension of the term in the action, 
with the scalar field itself having mass dimension $1$ and each derivative introducing
a mass dimension $M^1$.
Thus, 
the result in \rf{scalarLV2} is the $d=4$ limit and
the coefficients $k^\mn$ are dimensionless.
In general the coefficients $(k^{(d)})^{\mu\nu\la...} $  
have mass dimension $M^{4-d}$.

\subsubsection{Gravity sector case}
\label{gravity sector case}

The action from the gravity sector that includes both linearized Lorentz invariant 
and Lorentz-violating terms, 
can be described similarly to the scalar case.
However, 
with a multicomponent field, 
the details of the tensor algebra are more complicated.
First we note that 
linearized General Relativity can be derived from the  action
\beq
I_{GR} = -\frac {1}{4\ka} \int d^4x \, h_\mn G^\mn,
\label{GR}
\eeq
where the Einstein tensor is expressed in 
linearized form with terms of order $h^2$ and higher discarded.
Note that an action quadratic in $h_\mn$ yields field equations linear in $h_\mn$.

We now explain in some detail, 
the construction outlined in Ref.\ \cite{km16}.
The starting point for an action that generalizes \rf{GR} is
\beq
I =\frac {1}{8\ka} 
\int d^4x \, h_\mn \hat{K}^{(d)\mu\nu\rh\si} h_{\rh\si},
\label{gravaction}
\eeq
where $\hat{K}^{(d)\mu\nu\rho\sigma}$ is an operator given by
\beq
\hat{K}^{(d)\mu\nu\rh\si} = K^{(d)\mu\nu\rho\si \ep_1 ...\ep_{d-2}}\prt_{\epsilon_1}... \prt_{\ep_{d-2}}.
\label{operators}
\eeq
The operator contains partial derivatives that act on the gravitational field fluctuations $h_\mn$; 
the $K^{(d)\mu\nu\rho\sigma \ep_1 ...\ep_{d-2}}$ are a set of constants in the chosen coordinates. 
The mass dimension label $d$ refers to the natural units of mass that each term has. 
At this stage, 
the nature of these constants is unknown and in what follows
we explain the conditions applied to constrain them.

One derives the field equations via variation of the action with respect to the fields, 
similar to the scalar example above. 
Varying the action \rf{gravaction} with respect to the metric fluctuations $h_{\mu\nu}$, 
yields
\bea
    \delta I &=& \frac {1}{8\ka} \int d^4x \, [  
    \delta h_{\mu\nu}\, K^{(d)\mu\nu\rho\sigma\epsilon_1 ...\epsilon_{d-2}} \prt_{\epsilon_1}...\partial_{\epsilon_{d-2}} \,\,  h_{\rho\sigma}  
    \nonumber\\
    && \pt{\frac 18 \int d^4 x} 
    + h_{\mu\nu}\,
    K^{(d)\mu\nu\rho\si\ep_1 ...\ep_{d-2}} 
    \prt_{\ep_1}...\prt_{\ep_{d-2}} \, \de h_{\rh\si} ].
    \label{var1}
\eea
In order to completely factor out the variation of the metric field $\de h_\mn$, 
integration by parts is performed on the second term. 
(Note that in doing the integration by parts, 
we discard surface terms with derivatives of the fluctuations, 
which is a nontrivial step reflecting
the fact that the action contains an arbitrary number of derivatives, 
going beyond the usual first order derivative form of conventional dynamics.)
When $d$ is even, 
the integration by parts is done an even number of times, 
creating an overall positive value for the term; 
if $d$ is odd, the over term is negative in value. 
We can represent this with $(-1)^d$, 
and then obtain
\beq
    \de I 
     = \frac {1}{8\ka} \int d^4x \delta  h_{\al\be}  \left[K^{(d)(\al \be)(\mu\nu)\ep_1...\ep_{d-2}}+(-1)^d K^{(d)(\mu\nu)(\al\be)
    \ep_1...\ep_{d-2}}\right]\partial_{\ep_1}...\prt_{\ep_{d-2}}h_{\mu\nu}.
    \label{vary2}
\eeq
Since $h_{\mu\nu}$ is symmetric, 
we can indicate the symmetry with parenthesis in $\hat{K}^{(d)(\mu\nu)(\alpha\beta)}$.

There are two considerations in \rf{vary2} to investigate, 
the first being that only terms contributing to the field equations should survive. 
Thus we must have
\beq
    K^{(d)(\al\be)(\mu\nu)\ep_1...\ep_{d-2}}+(-1)^d \,K^{(d)(\mu\nu)(\al\be)\ep_1...\ep_{d-2}}\neq 0.
    \label{cond1}
\eeq
The second consideration is the imposition of the linearized gauge symmetry, 
i.e., 
$h_{\mu\nu}\rightarrow h_{\mu\nu}-\prt_{\mu}\xi_{\nu}-\prt_{\nu}\xi_{\mu}$, 
where $\xi^\mu$ is an arbitrary vector.\footnote{General gauge-breaking terms are considered in Ref.\ \cite{km18}.}
If we apply this transformation on the metric within the action \rf{gravaction}, 
i.e., $\de_{\xi}h_{\mu\nu}=-\prt_{\mu}\xi_{\nu}-\prt_{\nu}\xi_{\mu}$, 
we obtain, 
from \rf{vary2},
\bea
\delta_\xi I &=& \frac {1}{8\ka} \int d^4x \,  \prt_\al \xi_\be 
    \left[(-1)^d \hat{K}^{(d)(\mu\nu)(\al\be)}+\hat{K}^{(d)(\al \be)(\mu\nu)} \right] h_{\mu\nu},
    \nonumber\\
    &=& -\frac {1}{8\ka} \int d^4x \, \xi_{\nu} \left[(-1)^d \hat{K}^{(d)(\rh\si)(\mu\nu)}+\hat{K}^{(d)(\mu\nu)(\rh\si)} \right]\,\prt_{\mu}h_{\rh\si}.
    \label{diffvary}
\eea
Since $\xi_{\mu}$ is arbitrary and derivatives of $h_{\rho\sigma}$ are not necessarily zero, 
the second condition becomes
\beq
    \left[ (-1)^d \hat{K}^{(d)(\rho\si)(\mu\nu)}+\hat{K}^{(d)(\rh\si)(\mu\nu)} \right]\prt_{\mu}=0.
    \label{cond2}
\eeq

Under these two conditions \rf{cond1} and \rf{cond2}, 
there are three categories of coefficients.
These categories are based in part on discrete spacetime symmetry properties
of the terms in the action:
their behavior under CPT transformations, 
for which they can be even or odd.
Also, 
the possible tensor index symmetries categorize these coefficients \cite{km16}.
The three types of $\hat{K}^{(d)\mu\nu\rh\si}$ ``hat" operators are written as
\bea
    \hat{s}^{\mu\rh\nu\si}&=&s^{(d)\mu\rh\ep_1\nu\si\ep_2...\ep_{d-2}}\prt_{\ep_1}...\prt_{\ep_{d-2}}, \nonumber\\
    \hat{q}^{\mu\rh\nu\si}&=&q^{(d)\mu\rho\ep_1\nu\ep_2\si\ep_3...\ep_{d-2}}\prt_{\ep_1}...\prt_{\ep_{d-2}} \nonumber\\
    \hat{k}^{\mu\nu\rh\si}&=&k^{(d)\mu\ep_1\nu\ep_2\rh\ep_3\si\ep_4...\ep_{d-2}}\prt_{\ep_1}...\prt_{\ep_{d-2}}.
    \label{sqk}
\eea
The $\hat{s}$ operators have even CPT and mass dimension $d \geq 4$; $\hat{q}$ operators have odd CPT and mass dimension $d\geq 5$; $\hat{k}$ operators have even CPT and mass dimension $d\geq 6$.
The process also reproduces the GR terms.

The Lagrange density is then 
\bea
        \cL &=& \frac{1}{8\ka} \ep^{\mu\rh\al\ka}\ep^{\nu\si\be\la}\eta_{\ka\la}h_{\mu\nu}\prt_{\al}\prt_{\be}h_{\rh\si} 
        \nonumber\\
        &&+\frac{1}{8\ka} h_{\mu\nu}(\hat{s}^{\mu\rh\nu\si}+\hat{q}^{\mu\rh\nu\si}
        +\hat{k}^{\mu\rh\nu\si})h_{\rh\si},
         \label{gravlag}
\eea
where the first term is an equivalent way of writing the standard GR using the totally antisymmetric Levi-Civita tensor density $\ep^{\mu\rh\al\ka}$(equivalent to \rf{GR}). 
It should be remarked at this point that the Lagrange density in \rf{gravlag}
is the most general one constructed purely from the metric fluctuations $h_\mn$ and taken to quadratic order only.
While it includes only constant coefficients in \rf{sqk}, 
it maintains linearized gauge symmetry.
Terms in this Lagrange density can arise in spontaneous-symmetry breaking models, 
when the additional fluctuations (including possible Nambu-Goldstone and massive modes around the vacuum values have been ``integrated out" or ``de-coupled" \cite{bk06,abk10,seifert09,seifert18}.\footnote{Discussion of the SME framework including the fluctuations more generally can be found in Refs. \cite{kl21,b21}.}
On the other hand, examples exist
where the quadratic order Lagrange density in \rf{gravlag} can arise from models with explicit symmetry breaking. 
In either scenario, 
one is then left with an ``effective" Lagrange density, quadratic in the metric fluctuations around a flat background, 
in which the fluctuations do not appear. 
Proceeding, 
the resulting vacuum field equations from \rf{gravlag} are 
\beq
     0 = \, G^{\mu\nu} \, -[\frac{1}{4}(\hat{s}^{\mu\rh\nu\si}+\hat{s}^{\mu\si\nu\rh})
     +\frac{1}{2}\hat{k}^{\mu\nu\rh\si} 
     +\frac{1}{8}(\hat{q}^{\mu\rh\nu\si}
     +\hat{q}^{\nu\rh\mu\si}+\hat{q}^{\mu\si\nu\rh}+\hat{q}^{\nu\si\mu\rh})]\, h_{\rh\si}.
     \label{eom1}
\eeq

In the absence of Lorentz violation, the field equations \rf{eom1} reduce to $G^\mn=0$.
In the Lorentz gauge, this reduces to $\Box \bar{h}^\mn =0$, where
$\bar{h}^\mn = h^\mn - (1/2) \et^\mn h^\al_{\pt{\al}\al}$ and
$\prt_\mu \bar{h}^\mn=0$.
For plane wave solutions $\bar{h}_\mn = A_\mn e^{-i p_\al x^\al}$, 
this yields $p^2 = p^\al p_\al=0$. 
This provides the dispersion relation for GR, 
\beq
    \omega = |\vec p|,
    \label{dispGR}
\eeq
the equation of motion in energy-momentum space which describes the propagation for GWs. 
Using the residual gauge freedom in this limit, 
the number of independent components of the plane wave solutions can be reduced to $2$
and will take the form of \rf{TT} in the Transverse-Traceless gauge.

To find the dispersion relation for the modified equations \rf{eom1}, 
one again assumes a plane wave form above.
There are then at least two approaches to solving the
resulting equations, 
where the components of $h_\mn$ appear highly coupled 
with one another due to the extra symmetry-breaking terms in \rf{eom1}.
The equations \rf{eom1} retain the usual gauge freedom, 
and so one can proceed by choosing a gauge condition
and then decomposing the resulting equations into time and space components.
For example, 
using a temporal-type gauge $h_{0\mu}=0$
and a helicity basis for the spatial components, 
one can show that to first order in the coefficients for Lorentz violation,
still only $2$ degrees of freedom remain \cite{Mewes:2019}.
Alternatively, 
a gauge-independent method for deriving the dispersion relation that
uses differential forms exists \cite{km09}.

Despite the fact that only two physical propagating degrees of freedom 
remain in the leading order Lorentz violation case, 
the two modes generally travel at different speeds in the vacuum, 
resulting in birefringence,
and the frequencies of the modes are highly dispersive. (Note that in contrast, for the scalar field example in \eqref{disp3}, 
there is no birefringence effect because there is only 
one scalar mode whose propagation is modified.)\,
With a helicity basis choice of spatial coordinates, 
the two propagating modes can be shown to lie in the $+2$ and $-2$ helicity projections of the spatial components of the metric fluctuations $h^{ij}$.
The modified dispersion relation can be written as
\beq
    \omega = |\vec p| \, 
    \left( 1-\zeta^0 \pm |\vec{\ze}| \right),
    \label{dispEq}
\eeq
where
\beq
|\vec{\ze}|=\sqrt{(\zeta^1)^2 + (\zeta^2)^2 +(\zeta^3)^2}
\eeq
and
\bea
    \ze^0 &=& \frac{1}{4 |{\vec p}|^2} \left(-\hat{s}^{\mu\nu}\,_{\mu\nu}+\frac{1}{2}\hat{k}^{\mu\nu}\,_{\mu\nu}\right),
    \nonumber\\
    (\ze^1)^2+(\ze^2)^2 &=& \frac{1}{8 |{\vec p}|^4}
    \left(\hat{k}^{\mu\nu\rh\si} \hat{k}_{\mu\nu\rh\si}-\hat{k}^{\mu\rh}\,_{\nu\rh}\,\hat{k}_{\mu\si}\,^{\nu\si}+ \frac{1}{8}\hat{k}^{\mu\nu}\,_{\mu\nu}\,\hat{k}^{\rh\si}\,_{\rh\si} \right),
    \nonumber\\
    (\ze^3)^2 &=&\frac{1}{16|{\vec p}|^4}
    \left(-\frac{1}{2}\hat{q}^{\mu\rh\nu\si}\,\hat{q}_{\mu\rh\nu\si}
    -\hat{q}^{\mu\nu\rh\si}\,\hat{q}_{\mu\nu\rh\si}
    +(\hat{q}^{\mu\rh\nu} \,_{\rh}
    +\hat{q}^{\nu\rh\mu}\,_{\rh})\hat{q}_{\mu\si\nu}\,^{\si}   
    \right).
    \label{zetas}
\eea
All of the derivative factors $\prt_\mu$ from \rf{sqk} are replaced with momenta $\prt_\mu \rightarrow i p_\mu$.
The plus and minus signs indicate the different dispersion relations for each propagating mode, in vacuum (birefringence).
Note that the dispersion and birefringence effects depend on the arrival direction 
of the plane wave $\hat p$, 
revealing this to be a fundamentally anisotropic effect 
that differs from kinematical isotropic descriptions of symmetry breaking
\cite{myw12}.

\subsubsection{Gravitational wave signals}
\label{signals}

Since the terms involving the coefficients in \rf{dispEq} are already at leading order, 
they can be evaluated with the zeroth-order solution (e.g.,  $p^\mu = \om (1, \hat p) = |\vec p| (1, \hat p)$). 
This reveals that any effects associated with arriving plane waves should depend
on angular functions of the unit vector $\hat p$.
Further, 
since LIGO-Virgo analysis use angular sky map coordinate systems,
it is advantageous to use the machinery of spherical harmonics and spherical tensors.
We can decompose the above coefficients into spherical harmonic form, 
\bea
    \ze^0 &=& \sum\limits_{djm} \om^{d-4} \, 
    Y_{jm}(\hat{\textbf{n}})\,k^{(d)}_{(I)jm}, \label{spherical1}\\
    \ze^1 \mp i\,\ze^2 &=& \sum\limits_{djm} \omega^{d-4}\, _{\pm 4}Y_{jm}(\hat{\textbf{n}})\left(k^{(d)}_{(E)jm}\pm i k^{(d)}_{(B)jm}  \right), \label{spherical2}
    \\
    \ze^3 &=& \sum\limits_{djm} \om^{d-4} \, Y_{jm}(\hat{\textbf{n}})\,k^{(d)}_{(V)jm}.
    \label{spherical3}
\eea
In these expressions, $Y_{jm} (\hat{\textbf{n}}) $ are the usual spherical harmonics
with ${\hat n}=-{\hat p}$, 
while $_{\pm 4}Y_{jm}(\hat{\textbf{n}})$ are spin-weighted spherical harmonics.
The coefficients, 
formerly in cartesian tensor form in \rf{dispEq} are expressed
in spherical form $k^{(d)}_{(I)jm}$, 
$k^{(d)}_{(E)jm}$,
$k^{(d)}_{(B)jm}$,
and $k^{(d)}_{(V)jm}$,
where $j =0,1,...,d-2$ and $-j \leq m \leq j$.
The meaning of the subscripts $I,E,B,V$ and the 
relation between the two forms of the coefficients is determined by whether the terms are CPT odd or even and which mass dimensions they encompass, detailed in Refs.\ \cite{km09,km16,km17}.

In GR, there is no difference in the speed between gravitational wave polarizations; 
both travel at the speed of light (i.e., $v=\om / |\vec p| =1$). 
In the case of a Lorentz violation in the form in \rf{dispEq}, 
the speed of the waves is given by
\beq
v= 1-\zeta^0 \pm |\vec{\ze}|. 
\label{LVspeed}
\eeq
Given enough propagation distance from source to detector, 
a difference in arrival times may be detectable even for small Lorentz violation, 
a feature that has been used for photon tests 
of Lorentz invariance \cite{Kostelecky:2001mb,Kostelecky:2006ta,Kostelecky:2007zz,Kostelecky:2008be,Kostelecky:2013rv,Kislat:2017kbh,Friedman:2020bxa}. 
Using LVC data, 
we can test for these effects by  
looking for a phase deviation from GR via polarization comparisons. 
If Lorentz violation effects are not resolvable given current precision, 
we can then provide constraints for the LV coefficients. 

Modifications in the analysis code use the expressions for the gravitatonal wave strain polarizations. 
The plane wave solutions will have a phase shift $\de \psi_{\pm}$ due to terms in \rf{LVspeed} or \rf{dispEq}. Consider first the strain
\beq
h \sim e^{-i (\omega t -kl)}
\eeq
where $l$ is the distance travelled and $k$ is the wave number. The difference in phase grows in magnitude the farther 
the gravitational wave travels from the source to detectors. 
On cosmological scales, 
it is important to include effects on propagation time from the expanding universe using luminosity distance. Noting $k \sim |\vec{p}| = \omega / v$, inputting \rf{LVspeed}, and including distance and frequency alterations form cosmology, one finds the phase shift expression,
\beq
\de \ps_{\pm}=\omega_{obs}\int^z_0\,dz' \frac{(-\ze^0\pm |\vec{\ze}|)}{H(z')},
\eeq
where $H(z)$ is the Hubble parameter with redshift $z$ and the observed frequency is related to that emitted via $\omega_{obs}(1+z)=\omega_{emit}$.

For each mode, 
the modified phase shift can be written as
\beq
     \delta \psi_{\pm} =- \delta \pm \beta,
     \label{delpsi}
\eeq
where 
\begin{align}
 \de &= \om^{d-3} \tau \zeta^{(d)0}, \nonumber\\
 \be &= \om^{d-3} \tau |\vec{\zeta}^{(d)}|,\nonumber\\
 \ta &= \int_0^z dz \fr{(1+z)^{d-4}}{H(z)}
 \label{quant1}
\end{align}
and $ |\vec{\zeta}| = \om^{d-4}  |\vec{\zeta}^{(d)}|$ and $ \zeta^{0}= \om^{d-4}\zeta^{(d)0}$. The $\tau$ is the effective propagation time due to cosmological redshift $z$.   

It is useful to rewrite the coefficients in terms of effective angles $\vth$ and $\vph$ defined by 
\beq
\sin\,\vartheta=\frac{|\ze^1\mp i\ze^2|}{|\vec{\ze}|}, \pt{30} \cos\vartheta =\frac{\ze^3}{|\vec{\ze}|}, \pt{30} e^{\mp i \varphi}=\frac{\ze^1\mp i \ze^2}{\sqrt{(\ze^1)^2+(\ze^2)^2}}.
\eeq
Note that these angles are not the sky location angles $\th$, and $\ph$.
Also using the plus and cross polarizations \rf{TT}, 
the modified gravitational wave solutions in terms of the Lorentz-invariant solutions can be written
\bea
     h_{(+)} &=& e^{i\de} (\cos \be - i \sin \vth \cos \vph \sin \be )\, h^{LI}_{(+)} \nonumber\\
     && - e^{i\delta}\sin \be (\cos \vth + i \sin \vth \sin \vph )  \, h^{LI}_{(\times)} \nonumber\\
     h_{(\times)} &=& e^{i\de} (\cos \be +i \sin \vth \cos \vph \sin \be )\, h^{LI}_{(\times)} \nonumber\\
     && + e^{i\de}\sin \beta(\cos \vth - i \sin \vth \sin \vph )  \, h^{LI}_{(+)}.\label{h_LIV}
     \label{plcr}
\eea
The $h^{LI}_{(+,\times)}$ are the Lorentz-invariant gravitational wave for standard GR; one can retrieve GR as a limiting case as $\be \rightarrow 0$ and $\de \rightarrow 0$.

The measured signal at a given detector can be obtained from an equation of the form \rf{antenna}.
It is standard in the literature to adopt 
a Sun-centered Celestial-Equatorial coordinate system (or SCF frame)
for reporting measurements of the components of the coefficients 
for Lorentz violation either in the form $s^{TXY...},...$ 
or in spherical tensor form $k^{(d)}_{(I)10},...$ \cite{datatables}.
Under observer coordinate transformations, 
the coefficients transform as tensors.
In many cases, 
these transformations can be implemented as global Lorentz transformations
on the coefficients.
In the present case, 
we want to ensure the coefficients in the expression for the measured strain
are all expressed in terms of the SCF coefficients, 
thereby leaving any angular, sky location, dependence in the relevant angular variables.
Thus, when analyzing data, 
the signal generically will have extra angular, 
isotropy breaking, 
dependence on the sky angles.
This will differ significantly from the GR case.

\subsection{Unit changes and dimension}
\label{units}

For applications below, 
it becomes essential to convert from natural units to SI units when implementing
modifications into analysis code.
We note here several useful unit substitutions that can used for this and various key equations discussed previously.

Recall natural units are based on $\hbar = c =1$.
In these units, 
quantities can have dimensions of energy, typically expressed in 
terms of electron volts, as \rm{GeV}$=10^{9}$ \rm{eV}, for example.
For instance, 
mass dimension $d$ coefficients for Lorentz violation have units of
$M^{4-d}$.
To convert various quantities to SI units, 
we assume that the starting
action has units of joules meters $\rm{Jm}$.\footnote{Alternatively one can choose $\rm{Js}$ to match classical mechanics.}
For instance, 
the full Einstein Hilbert action in SI units can be written as
\beq
I_{EH} =  \frac{c^4}{16 \pi G} \int d^4 x \sqrt{-g} R,
\label{EH}
\eeq
or for the quadratic action limit of equation \rf{GR}, 
simply multiply by $c^4$.
Units of $\rm{kg\,m\,s^{-2}}$ \, come from the factor $\frac{c^4}{G}$, 
$\rm{m^4}$ \, from $d^4x$, 
and $\rm{m^{-2} }$ \, 
from the derivatives contained within the Einstein tensor. 
Implicit here is the assumption that the metric tensor $g_\mn$ is dimensionless (the Minkowski metric retains its form $\et_\mn={\rm diag} (-1,1,1,1)$).
Likewise, 
the Lorentz-violating action \eqref{gravaction} contains operators with SI units $\rm{m^{-2} }$ \, , 
and thus from \eqref{operators}, 
when introducing higher derivatives, 
the units of the coefficients compensate, 
thus the coefficients have units $\rm{m^{d-4} }$.

When converting the field equations \eqref{eom1} 
from position to momentum space, 
every partial derivative contributes 
a factor with Planck's constant, 
i.e., 
$\prt_\al \rightarrow \frac{i}{\hbar}p_\al$.
Schematically, 
the the position space equation has the form
\beq
\partial \partial\, h\,\, + s^{(4)}\,\partial\partial\,h\,\, +q^{(5)}\,\partial \partial \partial\,h \,\,+\,...=0,
\label{SIposition}
\eeq
where, e.g., for $d=4$ a term involving the $\hat{s}$ operators contains coefficients for $s^{(4)}$ coupled to two derivatives that act on $h$.  In momentum space,
\beq
(\frac{i}{\hbar})^2 pp\, h\,\,+ (\frac{i}{\hbar})^2s^{(4)}\,pp\,h\,\,
+ (\frac{i}{\hbar})^3q^{(5)}\,ppp\,h \,\,+ \,...=0,
\label{SImomentum}
\eeq
where the operators $\hat{s}$, $\hat{q}$, and $\hat{k}$ now contain $(\frac{i}{\hbar})^{(d-2)}\,p_{\alpha_1}...p_{\alpha_{d-2}}$ in place of partials. The units for the coefficients are unchanged, i.e., $\rm{m^{d-4} }$.

One must also keep track of the corrected time-component factors in the four-momenta, 
$p_{\alpha}=(-\frac{\hbar}{c}\om,\, \vec{p})$.  
For instance, 
the wave speed,
via the dispersion relation, 
becomes
\beq
v_{\pm}=\hbar\,\om/|\vec{p}|=c\left(1+c^2(-\ze^0 \pm |\vec{\ze}|\,\,)\right)
\label{vSI}
\eeq
where each $\ze$ quantity in \rf{zetas} inherits a factor of $(\frac{\hbar}{c})^2$.
To ensure the coefficients, $k^{(d)}_{(I)jm}$, $k^{(d)}_{(E)jm}$, $k^{(d)}_{(B)jm}$, and $k^{(d)}_{(V)jm}$ have SI units of $\rm{m^{d-4} }$, we redefine the equations \rf{spherical1}-\rf{spherical3} by implementing a factor of $c^{(2-d)}$, e.g.,
\beq
\ze^0= c^{(2-d)}\sum\limits_{djm} \om^{d-4} \, 
    Y_{jm}(\hat{\textbf{n}})\,k^{(d)}_{(I)jm},
    \label{zetaSI}
\eeq
with similar SI factors for $\ze^1, \ze^2, \ze^3$.

\section{Analysis method}
\label{lalsuite}

The coefficients for Lorentz and CPT violation can be measured from the comparison of the speed of gravitational and electromagnetic waves, an analysis that has been performed with gravitational-wave event \, GW170817 and the associated counterpart gamma-ray burst (GRB) \, GRB170817A to constrain coefficients of mass dimension 4 with improved accuracy~\cite{Abbott_2017}.
Using GW signals only, limits on mass dimensions 5 and 6 coefficients have been obtained from the non-observation of a delay between the arrival time of the $h_{+}$ and $h_{\times}$ polarizations in the LIGO and Virgo interferometers~\cite{km16,shao20,wang21}. 

The constraints on the birefringence parameters are obtained from the posterior samples inferred under the assumption of no symmetry breaking, and are limited by the detector resolution to determine the waveform peak frequency, focusing on information from signals at higher frequencies. We aim to compliment prior work by analysing directly the LVC interferometers strain in order to bypass the reliance on posterior parameters inferred under a GR model. Our analysis therefore fully takes into account the correlation between the SME coefficients and the source parameters, including dispersion or birefringence effects, during the inference process.

We have implemented the modification of the GW strain obtained in \rf{plcr} to estimate the coefficients for symmetry-breaking from the morphology of the signals.
As the dispersive and birefringent effects are degenerate with the source properties (e.g. the luminosity distance, due to the additional energy loss during the propagation) we perform a joint estimation of the source parameters and the coefficients for Lorentz and CPT violation taking into account the modifications at all frequencies of the waveform. 
We implement the Bayesian analysis into a version of the LIGO Algorithm Library suite \texttt{LALSuite} modified for our purposes as described below \cite{lalsuite}. 

\subsection{Implementation of the modified waveform}
\label{sec:lalsim}

The joint measurement of source and beyond-GR constraints have been performed for a variety of new physics parameterizations, including modification of the GW generation and propagation~\cite{Abbott:2020jks}.
Following a similar methodology, we  implement the modifications of the GW signals derived from the SME framework in the GW simulation package of \texttt{LALSuite}.
Such deformations can be anisotropic, as can be infered from the appearance of ${\hat n}$ in Eq.\ \rf{plcr} via $\be$ and $\de$.
Here we focus on the simplest coefficients that produce dispersion and birefringence via Lorentz and CPT violating effects,i.e., those of mass dimension 5. 
These coefficients are contained within $\beta$ in \rf{plcr} and obey the complex conjugate relation $k^{(d)*}_{jm}=(-1)^m k_{j(-m)}^{(d)}$, for $j=0,1,2,3$, $-j\leq m \leq j$. There are {\it a priori} independent coefficients in this set of terms \cite{Mewes:2019}. 
We display the first terms within $\beta$ in SI units:
\beq
\beta^{(5)}=\frac{\omega^2\tau^{(5)}}{2\sqrt{\pi}c} \,  \bigl\lvert k^{(5)}_{(V)00}- \sqrt{\frac{3}{2}}\sin \theta \left(e^{i\phi} \,k^{(5)}_{(V)11}+e^{-i\phi} \,k^{(5)*}_{(V)11}\right)+ \sqrt{3}\cos \theta\, k^{(5)}_{(V)10}+... \bigr\rvert.
\label{eq:beta_kv5}
\eeq
where the sky location of the source ($\th$, $\ph$) appears.
The coefficients are taken as expressed in the SCF in this expression.  

The general form of the signal observed in the interferometer is: \beq
S_A = F_{(+)} h_{(+)}+F_{(\times)}h_{(\times)}\label{strainang1},
\eeq
where $h_{(+,\times)}$ are the expressions \rf{plcr}, and  $F_{(+,\times)}$
are the (standard) detector response functions.
The expressions for $F_{(+,\times)}$ include the rotation angles relating different frames, e.g., the merger frame and the detectors frame as defined in the \texttt{LALSuite} software.

The effective propagation time $\ta$ parameter of equation \rf{eq:beta_kv5} is defined in
equation \rf{quant1} as an integral function of the redshift.
Since it needs to be evaluated for every value of the SME coefficients being tested, for computing time feasibility we instead probe the effective coefficient $(k^{(5)}_{(V)jm})_{eff} = \ta \ k^{(5)}_{(V)jm}$.
The value of the SME coefficient is recovered after convergence of the inference process further described in the following Section.

Finally we note that transformations of the coefficients under observer boosts are also computable. 
This would be important,  should it become necessary to include the motion 
of the Earth, the interferometers, or the motion of a source system's center of mass, relative to the SCF.
Currently, it appears the strain measurements are not sensitive to this level of nonrelativistic boosts (e.g. $v/c=10^{-4}$).

\subsection{Bayesian analysis}

After implementing the modification of the strain, we include the SME coefficients in \texttt{LALInference}, the parameter estimation package of \texttt{LALSuite}~\cite{Veitch:2014wba}.
\texttt{LALInference} performs Bayesian inference of the posterior probability of the GW source parameters with the inclusion of the systematic uncertainties due to the detectors resolutions.
The vector set of GR prior parameters,  $\vec{\theta}_{GR}$, includes intrinsic parameters describing the binary system (e.g. the black holes masses and spins) as well as extrinsic parameters placing it in the astrophysical environment (e.g. the sky location, distance, inclination).
We add to the preexisting parameters the SME coefficients $(k^{(5)}_{(V)jm})_{eff}$ described in Section~\ref{sec:lalsim} for the mass dimension 5 case, contained within $\vec{\theta}_{SME}$.

In order to include the correlation between the GR parameters and the SME coefficients, we perform a simultaneous inference of all the parameters, 
obtaining the joint posterior probability: 
\bea
   P(\vec{\th}_{GR},\vec{\th}_{SME}|d,I) = \frac{ P(d|\vec{\th}_{GR},\vec{\th}_{SME},I)\,\,P(\vec{\th}_{GR},\vec{\th}_{SME}|I)}{P(d|I)},
   \label{eq:bayes}
\eea
where $P(\vec{\theta}_{GR},\vec{\theta}_{SME}|d,I)$ is the posterior probability, $P(d|\vec{\theta}_{GR},\vec{\theta}_{SME},I)$ the likelihood, $P(\vec{\theta}_{GR},\vec{\theta}_{SME}|I)$ the prior probability and $P(d|I)$ the evidence, and any pertinent background information is included in $I$. 
We set a flat prior probability for $(k^{(5)}_{(V)jm})_{eff}$
bounded between $|(k^{(5)}_{(V)jm})_{eff}| \in [0 ; 10^{-10} ]$, with maximal value well above the existing constraints on the order of $10^{-15}$~\cite{shao20}.
The likelihood is computed in the frequency domain: 
\bea
   P(d|\vec{\th}_{GR},\vec{\th}_{SME},I) = \exp \sum_i \left[ - \frac{2 | \tilde{h}_i (\vec{\th}_{GR},\vec{\th}_{SME}) - \tilde{d}_i|^2 }{T  S_n(f_i)} - \frac{1}{2} \log \left( \frac{\pi T S_n(f_i)}{2} \right) \right], 
   \label{eq:llh}
\eea
where $\tilde{h}_i$ is the frequency-domain template signal, $\tilde{d}_i$ are the data observed by the interferometers, $T$ is the duration of the signal, and $S_n$ the power spectral density of the detector noise.

Due to the large number of parameters describing the GW emitted by the coalescence of binary systems, the posterior probability is infered with Markov Chain (MC) methods.
The chains perform semi-random walks in the parameter space where the recorded steps of the walks are proportional to the quantity in Eq.\ \rf{eq:bayes}. 
Different algorithms have been shown to be able to perform parameter inference, of which Markov Chain Monte-Carlo (MCMC) with parallel tempering and nested sampling are implemented in the LVC algorithm library. 
The method 
returns joint posterior probabilities of the GR parameters and the SME coefficients. 
From this we extract the marginalised posterior probability 
on a subset of parameters by integrating over the distribution of the other variables. 
The credible intervals are finally obtained by summing the volume of the posterior probability corresponding to the desired fraction of confidence. We present the results of Bayesian inference on simulated signals in Section~\ref{simulation}, and will provide the results of the ongoing analysis of LVC detections in a separate publication.

\section{Sensitivity study}
\label{simulation}

\begin{figure}[H]
    \centering
    \includegraphics[width=13cm]{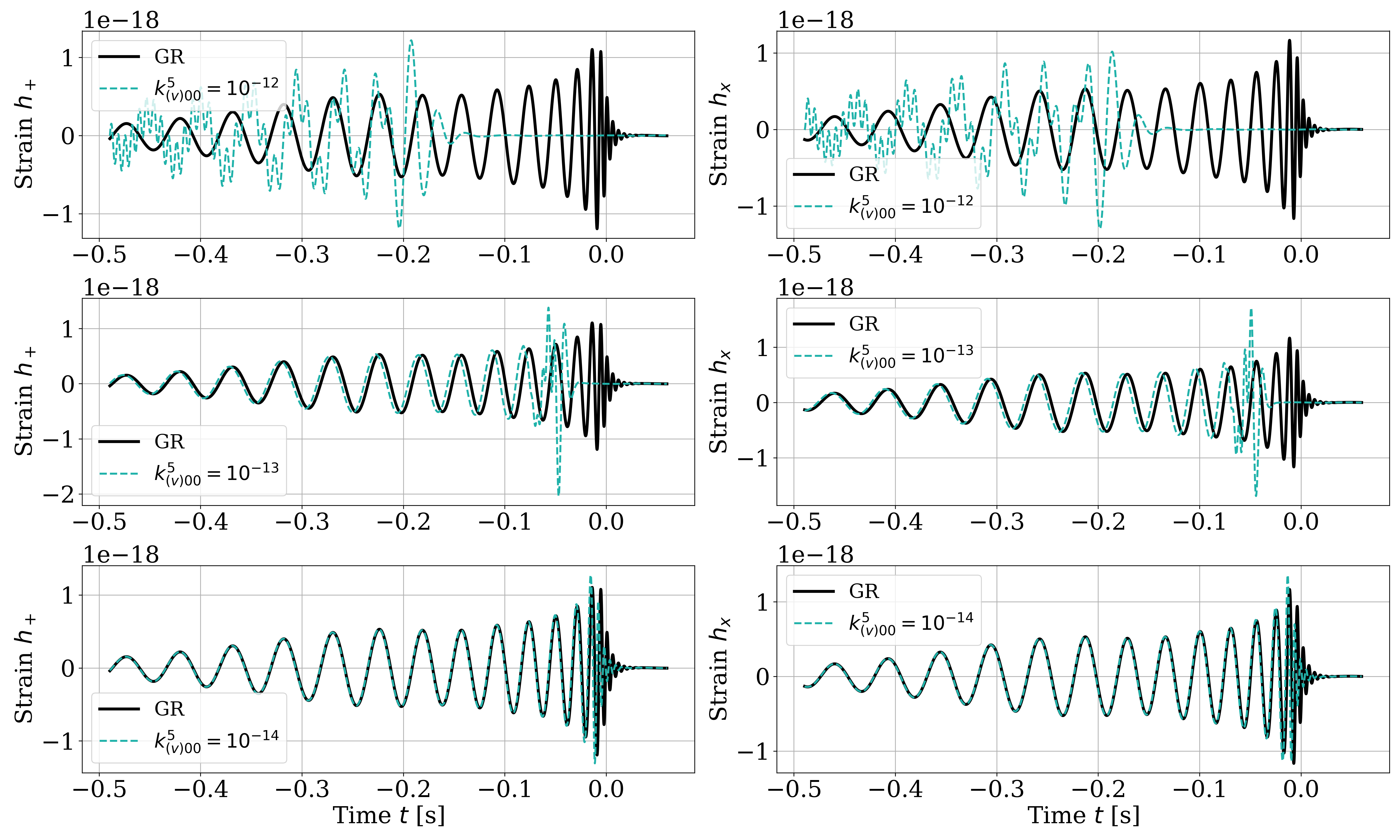}
    \caption{The above waveforms with varying $k^{(5)}_{(V)00}$ values, are for a simulated coalescence of a non-spinning binary system of black holes with $m_1=m_2=50 M_\odot$ located at a luminosity distance of 4 Gpc. GR denotes the case where $k^{(5)}_{(V)00} = 0$ and Lorentz violation is the case where $k^{(5)}_{(V)00}$ has the value specified above the plot.}
    \label{fig:kv5_00_sim}
\end{figure}

As an illustration,
we assume for the following, 
one non-zero coefficient $k^{(5)}_{(V)00}$ corresponding to isotropic polarization-dependent dispersion.
Figure \ref{fig:kv5_00_sim} plots the waveforms for both GR and the modified wave form for different values of $k^{(d)}_{(V)00}$.  
We assume a non-spinning binary system 
that has a luminosity distance of $4$ Gpc, 
and equal masses of $m_1=m_2=50 M_\odot$.
Note that significant differences in the waveform shape occur for coefficient values as small $10^{-13}$ m, impacting both the amplitude and frequency of the signal. 
This result can be compared with simulations using analytical template models presented in Ref.\ \cite{Mewes:2019}.
In the latter publication in Figures 1 and 2, 
simulated waveforms with non-zero coefficients for Lorentz and CPT violation appear to modify the waveform mostly around peak amplitude times, 
whereas the simulations here in Figure \ref{fig:kv5_00_sim} show modification at earlier times.

\begin{figure}[H]
  \centering
         \includegraphics[height=5cm]{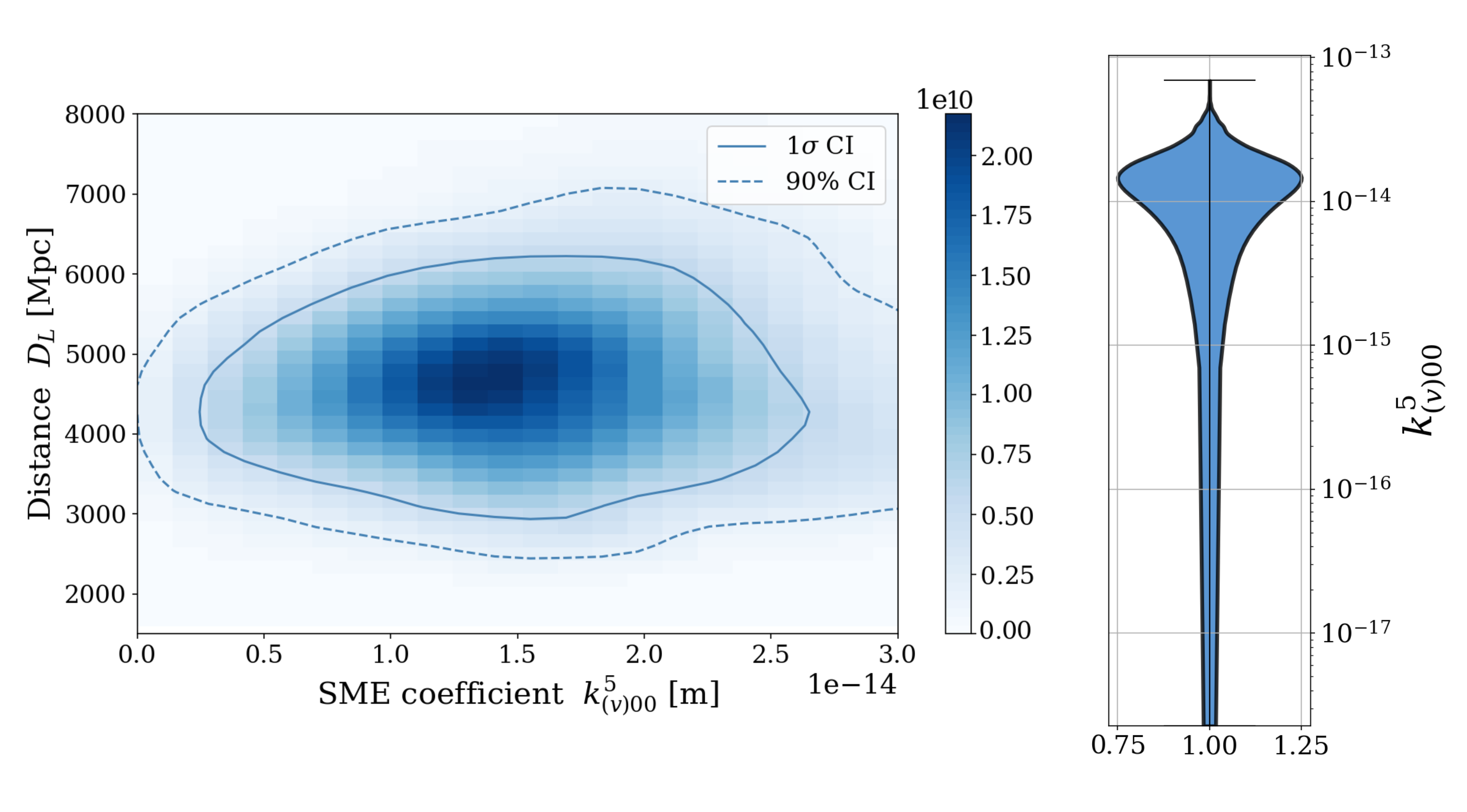}
\caption{Posterior probability density on the $k^{(5)}_{(V)00}$ coefficient for a simulated coalescence of a non-spinning binary system of black holes with $m_1=m_2=50 M_\odot$ located at a luminosity distance of 5 Gpc. The left figure shows the 1$\sigma$ and 90\% credible intervals in the $D_L - k^{(d)}_{(V)00}$ plane, the right figure shows the posterior probability of  $k^{(5)}_{(V)00}$ marginalising the source and systematical uncertainty parameters.
}
\label{fig:kv5_post}
\end{figure}

Using the methodology outlined in Section~\ref{sec:lalsim}, we perform a Bayesian inference of the source parameters and the coefficients for Lorentz-violation with simulated dispersed signals in order to study the potential to measure the coefficients with the LVC detections. 
We simulate a GW emitted by a non-spinning binary system of black holes with symmetric masses of $50 M_\odot$ located at 5 Gpc where the dispersion is controlled by one coefficient set to a value of $k^{(d)}_{(V)00} = 10^{-14}$.
Figure \ref{fig:kv5_post} shows the posterior probability on the luminosity distance and the coefficient, where both are recovered around the simulated values. 
The $1 \sigma$ credible interval shows a constraint on $k^{(d)}_{(V)00}$ where the zero value is excluded, showing that the coefficient can be measured with a single event providing that it is relatively large. 
The $k^{(d)}_{(V)00}$ posterior probability density marginalised over the source and systematic uncertainties is shown in the violin plot. 

These results, obtained with a single event, 
present encouraging prospects towards the measurement of the coefficients for symmetry-breaking with the current generation of GW interferometers. 
The second catalog of GW detections, 
encompassing the two first observing run as well the first half of the third observation run of the LVC, 
contains 50 events from the coalescence of binary systems of astrophysical compact objects, 
of which 46 are consistent with black hole systems~\cite{PhysRevX.11.021053}.
Comparing our results with measurements of the mass of the graviton, that also induce a modified dispersion of the GW signal, 
we note that the constraint has been improved of one order of magnitude from a single event to the analysis of a larger population of GW detections.
The constraint from GW159014 was $m_g \leq 1.2 \cdot 10^{-22} \textrm{ eV/c}^2$ while it is now $m_g \leq 1.76 \cdot 10^{-23} \textrm{ eV/c}^2$ when analysing 33 events from the second GW catalog~\cite{LIGOScientific:2016lio, Abbott:2020jks}.
Based on such results, we can conjecture that the constraints on the SME coefficients from the full catalog of GW detections will provide more stringent measurements than the preliminary sensitivity study shown here.
The robustness of such measurements due to the waveform modelling approximant has been explored in~\cite{LIGOScientific:2019fpa}, showing that those systematic uncertainties do not lead to a large bias nor re-estimation of the constraints at the current detector sensitivity.
Other studies show that transient noise may impact the measurement~\cite{Kwok:2021zny} by mimicking a GR deviation, an effect that we palliate by using the LVC-released power spectral densities and frequency ranges that exclude the presence of glitches in the strain data.  

\section{Conclusions and Future Work}
\label{conclusion}

We describe the implementation of an effective field theory framework for testing Lorentz and CPT symmetry into a version of the LIGO-Virgo Algorithm Library suite LALSuite. 
The Lorentz- and CPT-violating modifications include the coefficients controlling birefringence and dispersion effects on the gravitational wave polarizations.  
This work does not rely on posterior results inferred by the LVC that assume no deviations from standard GR; we implement the modifications due to dispersion directly at the level of the templates used for the Bayesian inference of the GW source and propagation parameters in order to incorporate the full information provided by the signal morphology.

Initially, one starts with the action in the effective field theory framework that is quadratic in the metric fluctuations $h_\mn$, \rf{gravaction}, 
and after theoretical constraints including gauge invariance,
we arrive at the general result in \rf{gravlag}.
From the field equations \rf{eom1}, 
a dispersion relation is derived, 
both in terms of the coefficients from the Lagrange density \eqref{dispEq}, 
and in terms of spherical coefficients in a special observer frame \eqref{spherical1}-\eqref{spherical3}. 
The result shows birefringence and dispersion for the two propagating modes; moreover these effects will vary with sky location of the source.
Then considering the expression for propagating 
and applying a modified phase shift, including cosmological considerations, one can rewrite the expressions for the plus and cross polarizations \eqref{plcr}, 
which are directly implemented within the modified \texttt{} package. 
Through Bayesian inference, we can perform a parameter estimation to constrain the coefficients for Lorentz violation 
Samples of visible effects are shown in the sensitivity plots in Section \ref{simulation}.

The theoretical derivations and sensitivity studies presented in this article precede the measurement of SME coefficients with the events detected 
by the LVC. 
This computationally intensive analysis is currently ongoing and the results will be reported in a future publication, where 
we aim to fulfil our analysis for coefficients for Lorentz and CPT violation 
of mass dimension five and six, 
with a global analysis.
In such a global analysis, 
the availability of what is now a plethora of GW sources across the sky has the potential to disentangle measurements for a large set of coefficients and thereby obtain an exhaustive search for signals of new physics.

\authorcontributions{Conceptualization, Q.G.B., K.O-A., L.H.; methodology, K.O-A. and L.H.; software, K.O-A., L.H.,T.D., and J.T.; validation, Q.G.B., K.O-A., L.H., T.D., and J.T.; formal analysis, Q.G.B., K.O-A., L.H., T.D., and J.T.; investigation, Q.G.B., K.O-A., L.H., T.D., and J.T.; resources, K.O-A., L.H., T.D., and J.T.; data curation, K.O-A., L.H., T.D., and J.T.; writing---original draft preparation, Q.G.B., K.O-A., L.H., and J.T.; writing---review and editing, Q.G.B., K.O-A., L.H., and J.T.; visualization, K.O-A., L.H., T.D.  All authors have read and agreed to the published version of the manuscript.}

\funding{This work was supported in part by the United states National Science Foundation (NSF) grants:  Q.G.B.\ and K.\ A.\ are supported by grant number 1806871 and J.T.\  is supported by grant number 1806990. L.H.\ is supported by the Swiss National Science Foundation grant 199307.  The author(s) would like to acknowledge the contribution of the COST Actions CA16104 and CA18108.  Computational resources were provided through the support of the NSF, STFC, INFN and CNRS, and LIGO Lab (CIT) supported by National Science Foundation Grants PHY-0757058 and PHY-0823459.}

\institutionalreview{Not applicable.}

\informedconsent{Not applicable.}

\acknowledgments{ 
The authors gratefully acknowledge the support of the NSF for the construction and operation of the LIGO Laboratory and Advanced LIGO as well as the Science and Technology Facilities Council (STFC) of the United Kingdom, the Max-Planck-Society (MPS), and the State of
Niedersachsen/Germany for support of the construction of Advanced LIGO and construction and operation of the GEO600 detector.  Additional support for Advanced LIGO was provided by the Australian Research Council, and further support from the Italian Istituto Nazionale di Fisica Nucleare (INFN), the French Centre National de la Recherche Scientifique (CNRS) and the Netherlands Organization for Scientific Research for the construction and operation of the Virgo detector and the creation and support of the EGO consortium.  The authors also thank two anonymous referees and Javier M.\ Antelis for valuable critiques of the manuscript.}

\conflictsofinterest{The authors declare no conflict of interest.}

\end{paracol}


\reftitle{References}

\externalbibliography{yes}
\bibliography{refs.bib}

\end{document}